13th International Conference on Topics in Astroparticle and Underground Physics, TAUP 2013

# The second-phase development of the China JinPing underground Laboratory

Jainmin Li[a], Xiangdong Ji[b,c], Wick Haxton[d,e] Joseph S.Y. Wang[e,*]

[a] *Tsinghua University. Beijing, China*
[b] *University of Maryland, College Park, MD, U.S.A.*
[c] *Shanghai Jiao Tong University, Shanghai, China*
[d] *University of California, Berkeley, CA, U.S.A.*
[e] *Lawrence Berkeley National Laboratory, Berkeley, CA, U.S.A.*

**Abstract**

During 2013-2015 an expansion of the China JinPing underground Laboratory (CJPL) will be undertaken along a main branch of a bypass tunnel in the JinPing tunnel complex. This second phase of CJPL will increase laboratory space to approximately 96,000 m$^3$, which can be compared to the existing CJPL-I volume of ∼ 4,000 m$^3$. One design configuration has eight additional hall spaces, each over 60 m long and approximately 12 m in width, with overburdens of about 2.4 km of rock, oriented parallel to and away from the main water transport and auto traffic tunnels. There are additional possibilities for further expansions at a nearby second bypass tunnel and along the entrance and exit branches of both bypass tunnels, potentially leading to an expanded CJPL comparable in size to Gran Sasso. Concurrent with the excavation activities, planning is underway for dark matter and other rare-event detectors, as well as for geophysics/engineering and other coupled multi-disciplinary sensors. In the town meeting on 8 September, 2013 at Asilomar, CA, associated with the 13$^{th}$ International Conference on Topics in Astroparticle and Underground Physics (TAUP), presentations and panel discussions addressed plans for one-ton expansions of the current CJPL germanium detector array of the China Darkmatter EXperiment (CDEX) collaboration and of the duel-phase xenon detector of the Panda-X collaboration, as well as possible new detector initiatives for dark matter studies, low-energy solar neutrino detection, neutrinoless double beta searches, and geoneutrinos. JinPing was also discussed as a site for a low-energy nuclear astrophysics accelerator. Geophysics/engineering opportunities include acoustic and micro-seismic monitoring of rock bursts during and after excavation, coupled-process *in situ* measurements, local, regional, and global monitoring of seismically induced radon emission, and electromagnetic signals. Additional ideas and projects will likely be developed in the next few years, driven by China's domestic needs and by international experiments requiring access to very great depths.

## 1. Introduction

On 8 September, 2013, at Asilomar, CA, the day before the 13[th] International Conference on Topics in Astroparticle and Underground Physics (TAUP), a town meeting was held to discuss plans for a very significant expansion of the existing China JinPing underground Laboratory (CJPL-I) as well as experiments that could benefit from the new laboratory spaces. This article is a written version of the town meeting summary that was presented during the TAUP meeting, in Session I of Underground Laboratories/Large Detectors.  It provides a synopsis of the town meeting's presentations and discussions, describing CJPL-II planning, some of the candidate physics projects that might be incorporated into the CJPL-II program, and opportunities for multidisciplinary geophysics and engineering studies during and after construction.

Section 2 of this summary describes the existing laboratory space/experiments and the planned expansion that will constitute CJPL-II. Section 3 describes the physics projects presented at the town meeting, including dark matter and other possible astroparticle experiments, neutrinoless double decay searches, and solar neutrino experiments. Proposed geophysics and other multi-disciplinary studies are described in Section 4.  Section 5 provides a summary. Throughout this paper we emphasize infrastructure aspects important to the science program.

## 2. China JinPing underground Laboratory and its planned extension

CJPL is currently the world's deepest underground laboratory with horizontal access. The vertical rock burden over the middle portions of the 17.5 km JinPing tunnels in Sichuan province, southwest China, averages about 2,400 m. The measured cosmic ray muon flux, $[2.0 \pm 0.4]$ x $10^{-10}$ / (cm$^2$-s), by Wu et al. (2013), corresponds to the water equivalent depth of 6,720 m (assuming a rock density of 2.8 g/m$^3$). The great depth makes CJPL an excellent location for dark matter and other rare-event searches that require very-low-background environments.

There are currently two dark matter search experiments at CJPL. The China Dark matter EXperiment (CDEX) uses point-contact germanium (PCGe) semi-conductor detectors described by Kang et al. (2013a, b) in a search for weakly interacting massive particles (WIMPs).  Liquid argon is used as the coolant. The goals of the current CDEX-1 1-kg PCGe detector, the upgraded 10-kg PCGe detector array of CDEX-10, and the future CDEX-1ton array include high sensitivity to low-mass WIMPS.  The goal is to achieve a nuclear recoil energy threshold of less than 300 eV. The second CJPL dark matter experiment is Panda-X, a liquid xenon dual-phase detector described by Gong et al. (2013) and Ni (2013). The initial stage with a 125-500 kg target (expected 25-300 kg fiducial mass) is proceeding in two steps during 2013-2014.  The long-term goal is to reach the ton scale. Panda-X deploys a cryogenic bus structure for cooling.

Both the current CDEX and Panda-X experiments, together with associated low background counting facilities and a data acquisition system, are located within CJPL Hall A, a ~ 2,000 m$^3$ room of approximate dimensions 40 m (length) x 6.5 m (width) x 6.5 m (height), see Fig. 1, from Yue and Wong (2013) and Li (2013a). The total volume of CJPL-I for physics is ~ 4,000 m$^3$, including access to the Hall A. JPL-I construction started in 2009.  The laboratory opened on 12 December 2010, as described by Li (2013a).

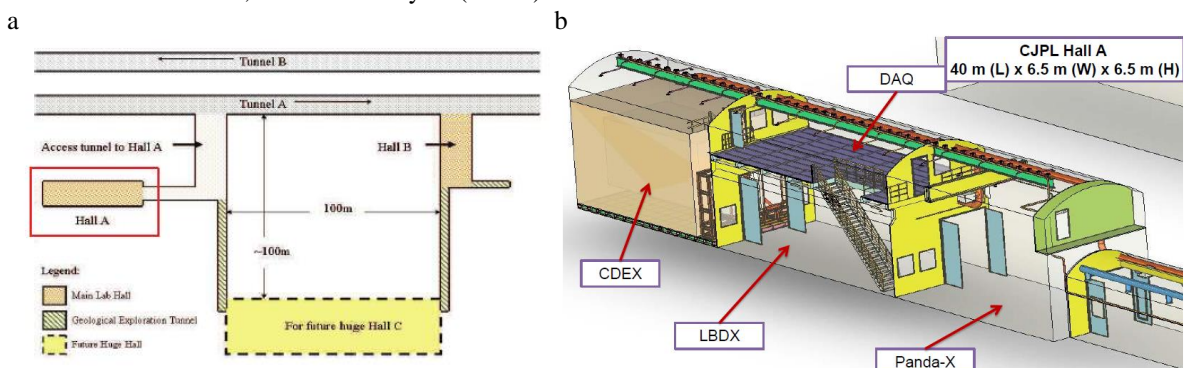

Fig. 1. Schematic diagram displaying features of CJPL-I: (a) Floor plan of Halls A (physics experiments) and B (rock mechanics studies), and an early proposal for future expansion (C);  (b) Layout of the CDEX and Panda-X  in Hall A and associated supporting equipment, from Yue and Wong (2013), Fig. 1, and Li (2013a), slide 55.

The Ya Long River in Sichuan province makes a big U-turn around JinPing mountain. The JinPing tunnels were excavated to connect two hydropower houses constructed into opposing sides of the mountain before and after the U-turn, in order to exploit the difference in river water levels, over the length of the tunnels, to generate electric power. The project includes two traffic tunnels, auxiliary tunnels A and B, shown in Fig. 1. Halls A (physics) and B (rock mechanics) of CJPL-I were constructed in the central portion of Tunnel A, as described by Chen (2012), Feng (2011), and Li et al. (2012a, b).

There are seven tunnels at the JinPing tunnel complex: four headrace tunnels, two traffic tunnels, and one water drainage tunnel, as shown in Fig. 2, from Feng (2011), Li et al. (2012a, b). The average tunnel length is 17.7 km. The excavation through the marble host rock was done initially by tunnel boring machine (TBM) and later by drill and blast (D&B). The main headrace tunnels have diameters as large as 13 m. The traffic tunnels are 6 m in diameter. Fig. 2b shows that the top of JinPing mountain is relatively flat over the central portions of the tunnels.

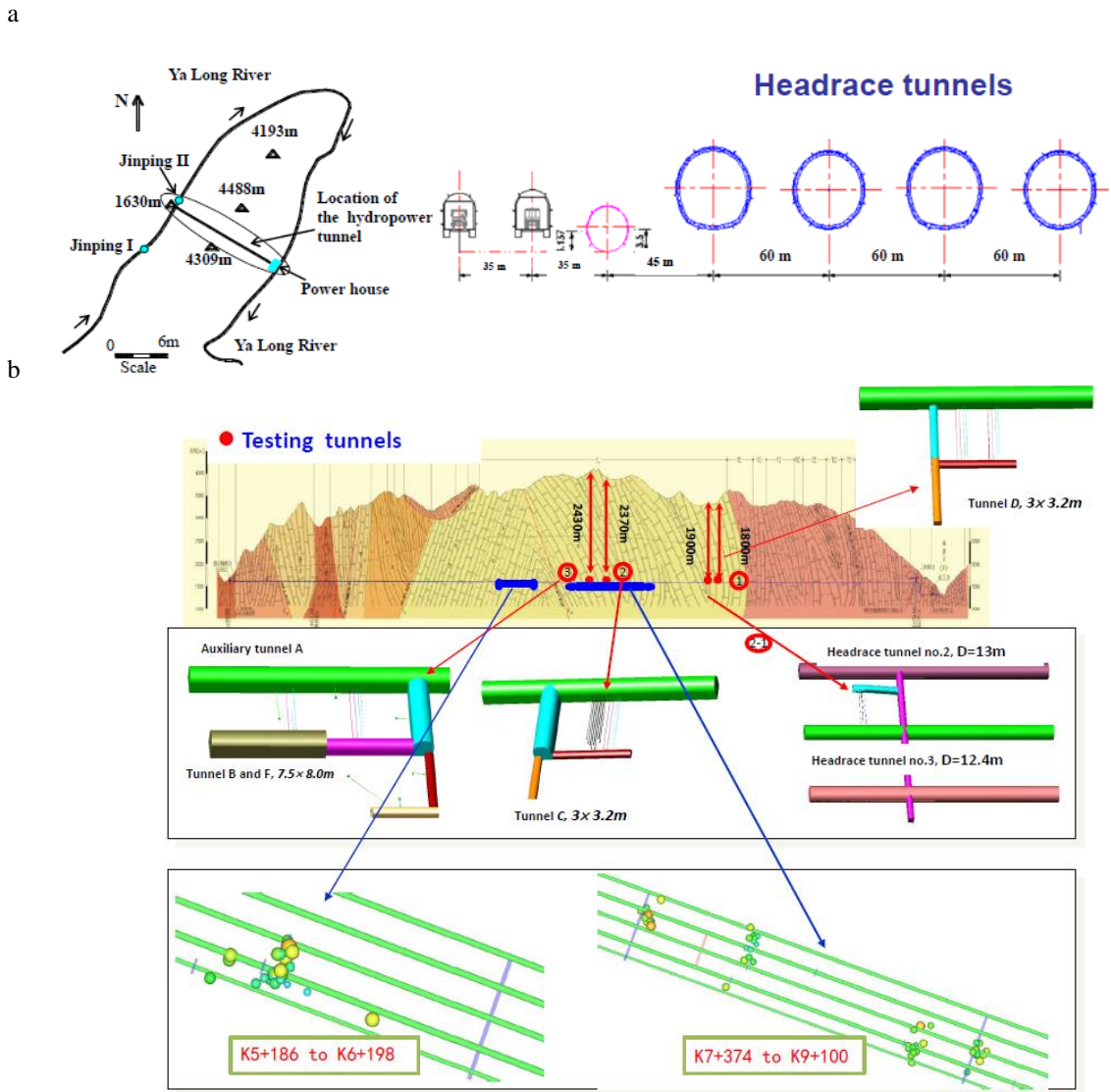

Fig. 2. Schematics of JinPing tunnel complex: (a) location and cross-sections of seven tunnels ; (b) additional small test tunnels located in the central portion of large diameter main tunnels (12.4 -13 m for the two headrace tunnels shown) and examples of micro-seismic events measured along the tunnels, from Feng (2011), slides 9 and 11.

The ultra-low value of muon flux mentioned in the first paragraph of this section agrees well with that obtained using the average depth in a flat overburden model as discussed by Wu et al. (2013). Fig. 2 will be further discussed in Sec. 4 (geoscience and rock mechanics studies). Marble rock samples collected in JinPing tunnels and in CJPL contain relatively low concentrations of K, Th, and radium. The measurement are $^{40}$K < 1.1 Bq/kg, $^{226}$Ra (609 keV) ~1.8 ± 0.2 Bq/kg, and $^{232}$Th (911 keV) < 0.27 Bq/kg, as reported by Li (2013a) and Chen (2012). The potassium and thorium upper bounds correspond to concentrations of less than 10 ppm and 67 ppb, respectively.

Dark matter and other rare event searches require very low background rates from penetrating muons and from radioactivity in the surrounding rocks. The radon concentrations in the passage or access tunnel and inside Hall A were measured with AlphaGuard PQ2000, yielding dose rates of 133 ± 24 Bq/m$^3$ in the LBF area sandwiched between CDEX and Panda-X laboratory spaces and 34 ± 7 Bq/m$^3$ inside CDEX. These values can be compared with radon concentrations at Laboratori Nazionali del Gran Sasso (LNGS) that were initially ~ 150 Bq/m$^3$, then were reduced to ~ 20 Bq/m$^3$ with ventilation improvements, as described by Plastino (2012). Radon measurements taken from February 2012 to July 2012 show eight or more periods when concentrations increased, for durations of up to a few days, as described by Li (2013b).

Preliminary studies are on ongoing at CJPL for correlations of the background measurements with earthquakes and other phenomena. Similar studies of correlations of radon temporal surges (as well as changes in other noble gases, uranium concentrations in water, etc.) with local and nearby earthquake events or other distributions have been carried out at LNGS and other underground laboratories, as discussed by Plastino (2012).

To accommodate next generation experiments that might require larger halls, lower background, and better shielding, CJPL-II will increase the available laboratory space near the existing CJPL-I location in the JinPing tunnel, as described by Li (2013a, b). The CDEX-1 ton and Panda-X-1 ton experiments would be mounted in the planned expansion spaces. The conceptual design is shown in Fig. 3. The current plan would create eight new ~ 60 m halls, excavated from Bypass Tunnel 1 of the traffic Tunnel A. Each of new laboratory halls would be 12 m (or wider) in width and 10 m (or taller) in height.

The total hall volume for CJPL-II would be approximately 96,000 m$^3$, as compared to 4,000 m$^3$ currently available at CJPL-I. Fig. 3b shows how CJPL-II would be oriented in relation to the seven tunnels of the JinPing tunnel complex. The schedule calls for the design of CJPL-II to be completed in 2013-2014 and for construction in 2015-2016, as described by Li (2013a, b). The expansion would provide the space needed for the 1-ton upgrades of CDEX and Panda-X, as well as significant additional space for other physics and earth science/multidisciplinary experiments, such as those discussed in Sec. 4.

This plan can be compared to the layout of LNGS, a laboratory with some structural similarities to CJPL: though less deep (1,400m peak rock overburden, yielding an effective depth of 3,000 mwe), LNGS is also a horizontal facility employing multi-purpose halls, each 20 m wide, 18 m high, and 100 m long for a total volume of ~ 180,000 m$^3$. As with CJPL, LNGS was created as part of a larger civil construction project, the 1987 boring of the Gran Sasso tunnel. The LNGS science program is quite broad and varied. In 2013 LNGS hosted six earth science related experiments: the topics include $^{222}$Rn - $^{14}$C - $^{3}$He monitoring of environmental radioactivity (ERMES), crustal deformation (GIGS), the maintenance of cryo-preserved biological materials in reduced-radiation environments (CRYO STEM), studies of aseismic creep strain episodes that might be associated with earthquakes (TELLUS), seismographic monitoring (UnderSeiS), and the effects of reduced-radioactivity environments on the biochemical behavior of living organisms (Pulex 2). The physics program includes five efforts on dark matter (CRESST, Dama/Libra, DARKSIDE, WARP, XENON), three on neutrinoless double beta decay (COBRA, CUORE, GERDA), two on solar or supernova neutrinos (Borexino, LVD), two connected with the CERN long-baseline neutrino program (ICARUS, OPERA), and one each on nuclear astrophysics (LUNA) and tests of the Pauli exclusion principle (VIP). Most of underground laboratories worldwide regard LNGS as a model for incubating and hosting major experiments and providing supporting infrastructure.

## 3. Potential physics experiments in CJPL-II

Neutrinoless double beta decay experiments, which have requirements similar to those of dark matter experiments such as CDEX and Panda-X, and are on a similar trajectory for scaling to and beyond one-ton, are underway at most underground laboratories. As the physics reach of such experiments depends on achieving exceptionally low background rates, the great depth of CJPL-II could be important to future efforts, particularly if accompanied by corresponding efforts to limit and mitigate backgrounds from natural radioactivity. Two configurations for the next generation Majorana/GERDA and nEXO experiments were discussed in the CJPL-II town meeting by Detwiler (2013). One of them deploys a compact linear shield and is likely a configuration suitable for CJPL-II, as shown by Detwiler (2013, slide 24). The other configuration requires wider and taller hall of a type that will be further discussed toward the end of this section. There was also a presentation from the Sino-German GDT Collaboration on the development of high purity germanium crystals for both double-beta decay and dark matter research as described by Majorovits (2013).

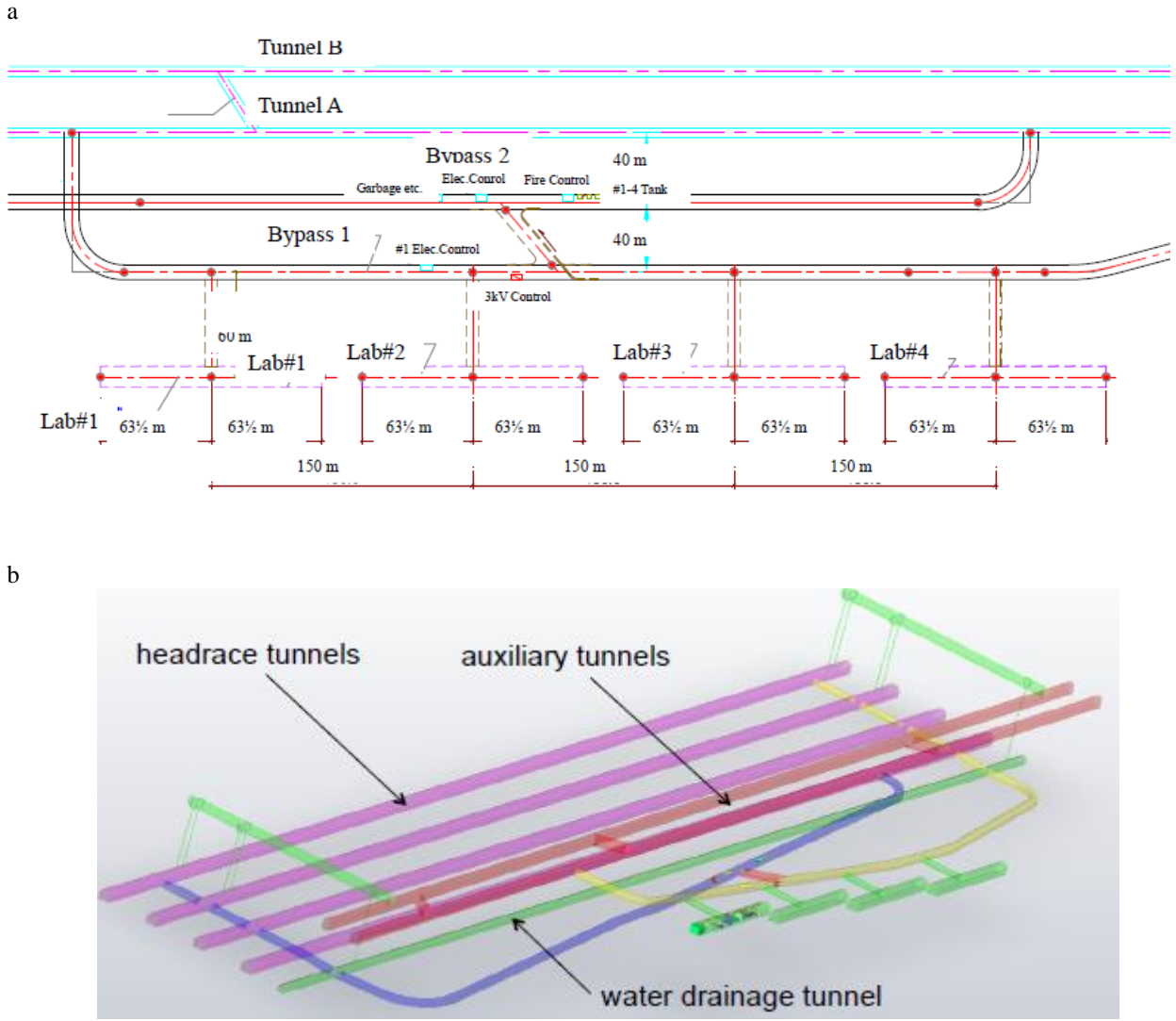

Fig. 3. Schematic diagram of CJPL-II: (a) Eight lab spaces connected by four 60 m long access drifts; (b) 3-D prospective of CJPL-II in relation to the JinPing tunnel complex, from Li (2013a) slides 79-80 or Li (2013b) slide 25.

The detection of the two carbon-nitrogen-cycle (CN) low-energy neutrino branches is one of the important challenges in solar neutrino physics. Such a measurement would provide a direct determination of the metalicity of the primordial gas cloud from which the solar system formed, a question now of great interest because of the >30% conflict between photospheric and helioseismic determinations of solar metalicity, as discussed by Orebi Gann (2013), slides 49, 59, 92 and 97. The CN neutrino fluxes can be measured in scintillation detectors similar to Borexino, provided the experiments are mounted at very great depths, to reduce the production of $^{11}$C when penetrating muons interact with $^{12}$C in the scintillator. CJPL and SNOLab may be the only sites with sufficient depth. This and other challenges in low-energy solar neutrinos, such as a precise determination of the pep flux, will likely require Borexino-like detectors with fiducial volumes of about one kton. CJPL-II could host such an experiment.

Nuclear reactions important to stellar evolution typically take place at energies far below the Coulomb barrier, where cross sections are very small. Laboratory experiments to measure these cross sections at energies relevant to astrophysics thus must contend with tiny rates, e.g., 1 event/month as shown by Lemut (2010, slide 2). DIANA is a U.S. proposal to construct a special-purpose low-energy accelerator for cross section measurements, building on the LNGS LUNA program. In Town Meeting discussions of DIANA, it was noted that the long halls of CJPL-II would work well for such an accelerator, particularly in a design using an inverted barrel.

The dark matter searches are a key activity in underground laboratories worldwide, as is apparent from the laboratory review by Bettini (2013). Current experiments typically employ detector masses ranging from a few to a few hundred kg, but generation-2 (G2) detectors at the 100kg to one-ton sale are anticipated by ~ 2017, while G3 multi-ton detectors are envisioned by ~ 2020, as discussed by D'Angelo (2013, slide 52). Many technologies are in use, including scintillating crystal detectors [e.g., Dama/Libra (NaI), KIMS (CsI)], cryogenic bolometers [CDMS (Si, Ge), EDELWEISS (Ge), CRESST (CaWO$_4$)], noble-gas liquid detectors [e.g., DarkSide (Ar), LUX (Xe), XENON (Xe), and DEAP/CLEAN (Ar,Ne)], and fluorine-based super-heated liquids [e.g., SIMPLE, COUPP and PICASSO]. The space requirements for G2 experiments are generally moderate. For example, an experiment like SIMPLE-IV requires ~ 1,200 m$^3$ to accommodate the envisioned 20-detector array, 2-m water shield, and data acquisition system, as described by Girard (2012). The background requirements for G3 experiments are daunting, given that the G3 goal of probing spin-independent cross section below $10^{-48}$ cm$^2$ corresponds to one event/ton/year. Cosmogenic neutrons are an important concern. While it has been argued that depths comparable to those of Sanford's Homestake Laboratory (4,200 mwe) are adequate given large, instrumented water shields and active scintillator shields, the additional factor-of-15 suppression of the muon flux that is achieved at CJPL depths allows for simpler designs, reducing costs while providing an addition margin of safety.

The improvement in $^{40}$K geoneutrino detection possible with CdWO$_4$ crystal detectors was discussed at the CJPL-II town meeting by Mei (2013). Geoneutrinos are produced by the decay of U, Th, and K in the earth's crust and core. By detecting these sources, one can determine the radiogenic contribution to heat flow, and test models of the chemical origin and composition of the earth's crust.

SNOLab was frequently referenced at the CJPL town meeting because of its similar depth and recent construction (opening in 2012). At 2000 m, SNOLab is the world second deepest underground laboratory. SNOLab has several specialized laboratory rooms, shaped to accommodate specific experiments.

SNOLab's cryo-pit space was extensively discussed in CJPL-II town meeting, in the context of the needs of G2 and G3 dark matter experiments as well as other low-counting rate experiments as described by Smith (2013), Sinclair (2013), and SNOLab (2006, Fig. 4.17 on page 45), and in searches for double beta decay as described by Detwiler (2013, slide 17). The cryo-pit is in the shape of a barrel, 18.3 m in diameter at the waist, 15.2 m in diameter at the base and top, and 19.8 m in height. A similar room was constructed for SNO, the solar neutrino experiment that preceded SNOLab. The SNO cavern is 22 m in diameter at its waist and 34 m in height, and is currently the site for SNO+, a double beta decay and solar neutrino detector currently under construction that will use Te-loaded scintillator. Similar barrel designs either exist or are being considered in other laboratories. The future ANDES laboratory along the Argentina-Chiles Agua Negra tunnel, which will provide an overburden of 1,750 m, has a large pit planned, 30 m in diameter and 42 m in height as described by Dib (2013). The existing Super-Kamiokande cavern, in Japan's Kamioka mine at depth of 1000 m, is 40 m in diameter and 55 m in height. Variations on the barrel design also are under consideration. CUPP in Finland, a laboratory that has been proposed in a mine that reaches a depth of 1,450 m, has plans for an ellipsoidal cavern to house the LBNO liquid argon detector of dimensions 64 m (w) by 51 m (h) by 103 m (l), as planned by Nuijten (2013). Hyper-Kamiokande, a proposed megaton water Cherenkov detector for long-baseline neutrino physics, atmospheric neutrino detection, and proton decay detection, has twin cylindrical caverns aligned horizontally and side-by-side, each 48 m wide, 54 m high, and 247.5 m long, as described by Yamatomi (2013) and Shiozawa (2013). Such a wide-span design is very challenging, even at the relatively moderate depths planned for Hyper-Kamiokande (~ 650m).

4**. Earth sciences, engineering, and multi-disciplinary studies**

Rock mechanics and geotechnical engineering studies were conducted in the tunnel complex and in test tunnels during JinPing excavation as described by Feng (2011) and Li et al. (2012a, b). As shown in Fig. 2, many micro-seimic events were observed. The micro-seismic and acoustic signals were used to determine, in the rock mass surrounding the tunnels, the extent of the zones disturbed by excavation. Excavation damaged zones were also identified by using a digital borehole camera to observe new fractures with openings > 2 mm. In addition to the disturbances accompanying excavation, intermediate and delayed rock burst events were observed in some sections along the tunnels. Intermediate rock bursts are associated with compressive failures, while delayed rock bursts are associated with tension and sheared failures. Micro-seismic events increased continuously, concentrating in the period during and directly after excavation, with a quiet period occurring before the delayed rock burst, see Feng (2011, slide 35). For example, the rock burst on 13 January 2011 occurred with a delay of 6 days 15 m away from the tunnel face, while the rock burst on 23 February 2011 occurred with a delay of 62 days 100 m away from the tunnel face. The tunnel cross sections were simulated to interpret the field observations on the locations of stress release zones and potential rock failure zones, as described by Feng (2011, slide 37). Under the high stress of ~ 70 MPa associated with maximum overburden of 2,525 m, the JinPing tunnels provide a unique opportunity to study rock burst events. The team of rock engineers will continue their investigations of rock bursts and associated phenomena starting in 2014 and continuing through the design and excavation stages of CJPL-II.

Injection of fluids can reduce the effective stress. Stress release ahead of the tunnel face can be exploited to enhance the rate of tunneling advance. Hydraulic fracturing operations have been developed in petroleum industry over decades and are widely applied, including shale gas development, as discussed by Dusseault (2013). Hydraulic fracturing experiments for mining were conducted in Australia in 2013 and are planned in mines of the Sudbury basin of Canada in 2014 and beyond. CJPL-II construction will provide an opportunity to explore hydraulic fracturing at great depths. Hydraulic fracturing operations have already been conducted at depths of ~2 km in, for example, Barnett shale in the U.S.A., as shown by Zoback et al. (2010). At great depths, the crust is critically stressed, with *in situ* permeabilities determined by fractures. At sufficient depths the brittle-ductile transition occurs, as shown by Zoback (2010, slide 30).

At depths over 2 km, it is generally believed that a transition occurs from the hydrostatic pressure regime to the lithostatic regime. Hydrostatic pressure is maintained by the continuous flow of water through flow paths, while at some depth the flow paths close under stress, so that isolated regions of high pressure may occur. This transition and other geophysics phenomena encountered at depths greater than 2 km were discussed by Wang and Li (2013).

Most underground laboratories have earth sciences and multi-disciplinary programs, in addition to physics. The LNGS program was described previously. In Kamioka a laser strain meter and a superconducting gravity meter were installed for geophysics studies. France's Modane Laboratory, Europe's deepest at a depth of 1,700 m, has plans to excavate a multidisciplinary underground sciences hall of 19 m (width) x 16 m (height) x 40 m (length) in an area adjacent to the existing laboratory and between the Modane highway tunnel and a new safety gallery that will be bored parallel to the highway, as described by Piquemal (2013, slide 19). The existing non-physics experiments at Modane include the use of gamma spectroscopy for biological studies and $^{137}$Cs wine dating. In Spain's Canfranc Laboratory, at a depth of 850 m, the GEODYN earth sciences observatory uses seismographs and laser strain meters mounted on ground surfaces and within the tunnel to monitor the entire geodynamical spectrum, as described by Diaz (2012). In the U.K.'s Boulby Laboratory, at a depth of 1,100 m, the Palmer Laboratory facility (where ZEPLIN and other physics experiments were located) has been converted into the International Subsurface Astrobiology Laboratory, where studies of the microbiology of deep subsurface salts are conducted, as described by Cockell et al. (2012). The ANDES Laboratory in Argentina-Chile plans to link seismograph networks in the region, as described by Dib (2013).

The 500-m-deep LSBB in France focuses on inter-disciplinary studies. In addition to the SIMPLE dark matter search, the LSBB hosts a "capsule" 8 m in diameter, 28 m long, shielded by a 2 cm-thick steel wall, resting on shock absorbers, and surrounded by a 2 m thick reinforced concrete as described by Waysand (2006). A superconducting quantum interferometer [SQUID]$^2$ device was mounted within this capsule. Its operations are limited only by fluctuations in the earth's magnetic field, as discussed by Waysand et al. (2009). One hour before the 2008 Sichuan-Wenchuan M8.1 earthquake, recurrent and well-defined ionospheric long-wavelength magnetic signals were detected by Waysand et al. (2010). A French-South African collaboration is now using both the low temperature [SQUID]$^2$ in France and the high temperature SQUID in South Africa to look for correlated global magnetic signals.

The participation of investigators from CJPL-II and/or ANDES in such studies would extend the global network of magnetic monitors for earthquake-related signals.

## 5. Discussion and summary

The construction of CJPL-II will provide laboratory space of ~ 96,000 m$^3$, an over twenty-fold increase over the space available in CJPL-I. The great depth, by reducing the muon flux to unprecedented levels, will simplify the design and construction of next generation dark matter, neutrinoless double-beta decay, and solar neutrino detectors. CJPL-II room design and the horizontal access could be important to projects like DIANA, a nuclear astrophysics facility that is likely to be constructed and tested on the surface, before being disassembled and transported to its permanent underground location. The lab's great depth is important to many earth sciences, engineering, and multi-disciplinary studies, where extreme conditions can help test our understanding.

The town meeting participants made several helpful suggestions. In addition to the barrel-shaped cryo-pit discussion, the participants advocated for a design that would provide room connections to two access drifts, for improved safety. Neighboring laboratory spaces could then be linked by an additional drift, providing each room with routes to two access drifts, useful in the event that an emergency blocks one of them.

The horizontal-access CJPL-I is the world's deepest underground laboratory. The additional space that will be available when CJPL-II is finished – equivalent to about half of that now available at LNGS, the world's largest laboratory – will accommodate a significant program of physics and multi-disciplinary studies. Further expansion is possible, if the space in CJPL-II someday proves inadequate. The CJPL-II site is 60 m from Bypass Tunnel 1, as shown in Fig. 3. A second bypass tunnel nearby could be utilized to further expand CJPL-II. Earth science studies will likely include locations way from the main laboratory, along sites of particular interest within the JinPing tunnels.

With CJPL-II design occurring in 2013-2014 and construction scheduled for 2015-2016, there is a need for international involvement now. Further meetings in the style of the Asilomar CJPL-II town meeting will be scheduled. Slides from town meeting presentations are available on the TAUP2013 web site.